# Stratified B-trees and versioning dictionaries.

Andy Twigg[*], Andrew Byde[*], Grzegorz Miłoś[*], Tim Moreton[*], John Wilkes[†*] and Tom Wilkie[*]
[*]*Acunu*, [†]*Google*
firstname@acunu.com

## Abstract

A classic versioned data structure in storage and computer science is the copy-on-write (CoW) B-tree – it underlies many of today's file systems and databases, including WAFL, ZFS, Btrfs and more. Unfortunately, it doesn't inherit the B-tree's optimality properties; it has poor space utilization, cannot offer fast updates, and relies on random IO to scale. Yet, nothing better has been developed since. We describe the 'stratified B-tree', which beats the CoW B-tree in every way. In particular, it is the first versioned dictionary to achieve optimal tradeoffs between space, query and update performance.

## 1 Introduction

The B-tree was presented in 1972 [1], and it survives because it has many desirable properties; in particular, it uses optimal space, and offers point queries in optimal $O(\log_B N)$ IOs[1]. In 1986, Driscoll et al. [7] presented the 'path-copying' technique to make pointer-based internal-memory data structures fully-versioned (fully-persistent). Applying this technique to the B-tree, the CoW B-tree was first deployed in the Episode File System in 1992. Since then, it has been the underlying data structure in many file systems and databases, for example WAFL [10], ZFS [4], Btrfs [8], and many more.

Unfortunately, the CoW B-tree does not share the same optimality properties as the B-tree; in particular, every update may require a walk down the tree (requiring random reads) and then writing out a new path, copying the previous blocks. Many file systems embed the CoW B-tree into an append-only log file system, in an attempt to make the writes sequential. In conjunction with the garbage cleaning needed for log file systems, this leads to large space blowups, inefficient caching, and poor performance. Until recently, no other data structure has been known that offers an optimal tradeoff between space, query and update performance.

This paper presents some recent results on new constructions for B-trees that go beyond copy-on-write, that we call 'stratified B-trees'. They solve two open problems: Firstly. they offer a fully-versioned B-tree with optimal space and the same lookup time as the CoW B-tree. Secondly, they are the first to offer other points on the Pareto optimal query/update tradeoff curve, and in particular, our structures offer fully-versioned updates in $o(1)$ IOs, while using linear space. Experimental results indicate 100,000s updates/s on a large SATA disk, two orders of magnitude faster than a CoW B-tree.

Since stratified B-trees subsume CoW B-trees (and indeed all other known versioned external-memory dictionaries), we believe there is no longer a good reason to use the latter for versioned data stores. Acunu is developing a commercial in-kernel implementation of stratified B-trees, which we hope to release soon.

## 2 Versioned dictionaries

A versioned dictionary stores keys and their values with an associated version tree, and supports the following operations:

- `update(key, value, version)`: associate a value to the key in the specified leaf version

- `range_query(start, end, version)`: return every key in the range [start,end] together with the value written in the closest ancestor to the specified version

- `clone(version)`: create a new version as a child of the specified version[2]

- `delete(version)`: delete a given version, and free the space used by all keys written there and no longer accessible by any version.

A versioned dictionary can be thought of as efficiently implementing the union of many dictionaries: the *live* keys at version v are the union of all the keys in ancestor versions, where if a key appears more than once, its closest ancestor takes precedence. If the structure supports arbitrary version trees, then we call it (fully-)versioned;

---

[1]We use the standard notation $B$ to denote the block size, and $N$ the total number of elements inserted

[2]Sometimes the literature refers to a 'snapshot' operation – here, a snapshot is exactly equal to cloning a leaf node.

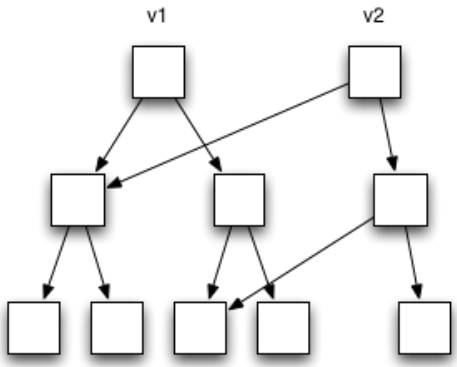

Figure 1: A copy-on-write B-tree

if it supports only linear version trees, we call it partially-versioned. We are interested in data structures that have optimal tradeoffs between space, and the performance of the operations above.

## 3 Copy-on-Write B-trees

The basic idea (see Figure 1) is to have a B-tree with many roots, one for each version. Nodes in this B-tree are versioned, and updates can only be done in a node of the same version as the update. When starting an update in version $v2$, we first ensure that there is a suitable root associated to $v2$ (by duplicating the root for $v1$, the parent of $v2$, if necessary), then follow pointers in the same way that one would in a normal B-tree; every time a node is encountered whose version is other than $v2$, it is copied to make a new node of version $v2$, and the parent node's pointer is updated to point to the new node, before the update continues down the tree. A lookup proceeds as in a B-tree, starting from the appropriate root.

Over time, variants of the CoW B-tree have appeared, notably in ZFS, WAFL, Btrfs, and more. Rodeh [14] provides a detailed discussion of variations of copy-on-write (shadowing) techniques, along with their pros and cons. The CoW B-tree has three major problems:

**Space efficiency:** each update may cause an entire new path to be written – to update a 16-byte key/value pair in a tree of depth 3 with 256K block size, one must first do 3x256K random reads and then write 768K of data (see http://bit.ly/gL7AQ1 which reported 2GB of log data per 50MB of index data).

**Slow updates:** each update may require a set of random reads (reading the old path) and a set of writes. Random reads limit prefetching ability (on SSDs, and are simply terrible on hard drives). Some file systems attempt to make the writes sequential by embedding the CoW tree in an append-only log, along with some way of garbage-collecting nodes no longer accessible, thus avoiding so-called 'in-place' updates (see, e.g. ZFS[3] and WAFL[4]). This approach has two fundamental problems: firstly, the excessive space usage inherent in CoW means that these writes present a lot of future work for the garbage collector; at 50x space blowup, the garbage collector has to work 50x harder to keep ahead of the input stream. Secondly, little is known in theory about guarantees for garbage collection in log file systems, particularly when the system does not experience idle time.

**Reliance on random IO:** Like the B-tree, the CoW B-tree relies heavily on random IO to scale, both for random reads and writes. Over time, the leaves of the tree tend to be scattered randomly across disk, which causes problems for efficient garbage collection and for efficient prefetching.

Therefore, we set out to answer the following open questions:

- Can we achieve the same query/update bound as the CoW B-tree, but with linear space?

- Can we achieve much better update bounds if we relax the query bound, using linear space?

- Can we scale with sequential IO rather than random IO?

In the unversioned case, the first two questions were answered in the affirmative by Brodal et al.[5], and the third by Bender et al. [3]. In particular, they showed that a small reduction in query performance could yield a huge improvement in update cost. Until now, no similar story was known for the versioned case. The space problem was tackled by Becker [2] who developed the multiversion B-tree (MVBT); however, it only offers slow updates and is only partially versioned. Lanka et al. [12] developed two fully-persistent B-tree variants, but for both variants, either there is a large space blowup, or range queries may be far from optimal, and hence don't advance the state-of-art beyond CoW and MVBT. Table 3 summarises these points; we use $N_v$ to count the total number of keys *live* (accessible) in version $v$, as opposed to $N \geq N_v$, the total number of elements written in all versions.

## 4 Multiversion B-trees

The multi-version B-tree (MVBT) was introduced by Becker [2]. It offers the same query/update bounds as the CoW B-tree but only requires asymptotically optimal $O(N)$ space. As previously mentioned, the MVBT only allows partial versioning – linear version trees; one way of viewing our stratified B-tree constructions is as a fully-versioned MVBT that also offers tunable query/update tradeoffs.

---

[3]http://blogs.sun.com/bonwick/entry/space_maps
[4]http://blogs.netapp.com/extensible_netapp/2009/04/understanding-wafl-performance-the-f-word.html

| Structure | Versioning | Update | Range query (size $Z$) | Space |
|---|---|---|---|---|
| B-tree [1] | None | $O(\log_B N)$ reads $O(\log_B N_v)$ writes | $O(Z/B)$ random IOs | $O(N)$ |
| Append-only CoW B-tree [ZFS] | Full | $O(\log_B N_v)$ reads $O(\log_B N_v)$ writes | $O(Z/B)$ random IOs | $O(NB \log_B N)$ |
| MVBT [2] | Partial | $O(\log_B N_v)$ reads $O(1)$ writes | $O(Z/B)$ random IOs | $O(N)$ |
| Stratified B-tree[this paper] | Full | $O((\log N_v)/B)$ reads $O((\log N_v)/B)$ writes | $O(Z/B)$ sequential IOs | $O(N)$ |

Table 1: Comparing the cost of basic operations on versioned dictionaries. $N$ is the total number of keys in the system, and $N_v$ is the number of keys *live* at version $v$.

The basic idea underlying the MVBT is to extend the traditional B-tree with *versioned pointers* inside nodes – each pointer additionally stores two integers, describing the set of (totally-ordered) versions for which the pointer is *live*. To query the tree for a specific version $v$, we first find the correct root node (with an additional $O(\log_B |V|)$ IOs), then at every node, extract the set of live pointers for $v$ and treat these as an unversioned B-tree node. Updates are more complex and require an additional version split operation in addition to the unversioned key split operation.

## 5 Log-structured CoW B-trees

In order to make the random IOs necessary for an update to a CoW B-tree sequential, a popular approach is to implement it within a Log File System (LFS), e.g., as first presented by Rosenblum and Ousterhout [15]. The main advantage of log file systems is that they offer the potential to transform random writes into large sequential IOs. The achilles heel of a LFS is *cleaning* – recovering invalidated (e.g. overwritten) blocks in order to reclaim sufficiently large contiguous regions of freespace.

Soules et al. [17] compare the metadata efficiency of a versioning file system using both CoW B-trees and a structure (CVFS) based on the MVBT. They find that, in many cases, the size of the CoW metadata index exceeds the dataset size. For example, in one trace, the versioned data occupies 123GB, yet the CoW metadata requires 152GB while the CVFS metadata requires 4GB, a saving of 97%.

**Analysis.** The basic analysis from [15] is the following: assume that during cleaning, we recover $N$ segments of average utilisation $\mu$. Then the write amplification $\rho$ due to the LFS is

$$\rho = \frac{N + \mu N + (1-\mu)N}{(1-\mu)N} = \frac{2}{1-\mu}.$$

The CoW B-tree requires $O(\log_B N_v)$ random reads and $O(\log_B N_v)$ blocks to be written, so the cost of a write is $O(\log_B N_v(1+\rho))$, roughly a factor $O(B(1+\rho)) \gg 1$ slower than the stratified B-tree. In practice, for a 512KB block with 128-byte entries, $B \approx 4096$.

One major difficulty of log file systems is guaranteeing that the write amplification $\rho$ will be small. Rosenblum and Ousterhout [15] reported that it depends heavily on the total storage utilisation $s$. Under uniform rewrites, greedily choosing segments with lowest utilisation gives $\rho \approx 6$ when $s \approx 0.8$. Robinson [13] derived analytic fluid limits for $\mu$ under the same conditions and showed that, for $s = 0.8, \rho \approx 5.4$. Under access patterns exhibiting locality, a different strategy does better by trying to force segment utilisations into a bimodal distribution (recently-written data is likely to be overwritten soon, so hot segments should be cleaned at a lower utilization than cold segments).

### 5.1 Solid State Drives

SSDs appear to offer a new lease of life to the LFS-structured CoW B-tree. SSDs handle small random reads very effectively, but perform poorly at random writes (the recent Intel X25M [11] can perform 35,000 4KB random reads/s, but an order of magnitude fewer writes). However, they can perform very efficiently for large sequential writes. Thus they appear to be well-suited to the LFS CoW B-tree that performs random reads and relatively large sequential writes. Stratified B-trees also only use large sequential writes for updates (and indeed, no random reads). Therefore, although the CoW B-tree can perform significantly better for updates on SSDs than disks, it gains no advantage over the stratified B-tree. Assuming blocks of size 4KB and 128 byte entries, $B \approx 16$, the CoW B-tree is roughly a factor $16(1+\rho) \approx 96$ times slower than the stratified B-tree.

# 6 Stratified B-trees

Stratified B-trees dominate CoW B-trees (with or without append-only logs). They can be written without append-only logs and heuristic-based garbage collectors. They require asymptotically optimal $O(N)$ space, offer an optimal range of tradeoffs between updates and queries, and can generally avoid performing random IO. In particular, one construction offers updates two orders of magnitude faster than CoW B-trees, with a small slowdown in point queries. However, for range queries, it can generally perform a large amount of sequential IO, allowing it to perform substantially better for these types of queries.

## 6.1 The Idea

Stratified B-trees are, like B-trees, quite simple to describe at a high level. The difficulty lies in proving that they have the desired properties. Here we describe in outline one of the main stratified B-tree constructions, known as the Stratified Doubling Array (SDA). Proofs and details of the algorithms can be found in [6].

The high-level structure is similar to that of the COLA of Bender et al. [3] - we store a collection of arrays of sorted (key-version-value) tuples, arranged into levels, with 'forward pointers' between arrays to facilitate searching. Each array $A$ is tagged with a set of versions $W$, so we write $(A, W)$. Arrays in level $l$ are roughly twice as large as arrays in level $l - 1$, hence *doubling*, and arrays in the same level typically have disjoint sets of versions associated to them, hence *stratified* in version space.

**Insertions, promotions and merging.** Inserts go first to an in-memory buffer, then are flushed in sorted order to the lowest level. When there are multiple arrays in the same level with intersecting version sets, we merge them all together. The resulting array may be too large to exist at the current level, in which case part of it may be extracted and *promoted* to the next level. Deletions are handled by inserting 'tombstone' elements that remove keys at the given version when they are encountered during merges.

**Density.** The notion of *density* is critical - we say an array $(A, W)$ has density $\delta$ if, for every version $w$ in $W$, at least a fraction $\delta$ of the elements in $A$ are *live* at $w$. The key to achieving performance is to ensure that density is not too low, while the key to achieving good space bounds is not to require it to be *too* high. For example, if we insist on density 1, then every array can only contain elements for a single version, which may result in large space blowup and potentially high update cost in the worst case. On the other hand if we allow very low density, then clearly we have space $O(N)$ but there is no guarantee on lookup or range query performance. Fortunately it is possible to strike a balance between these two extremes – it turns out that a lower bound of $1/3$ works well, and we can maintain this density balance without blowing up space or incurring a performance overhead.

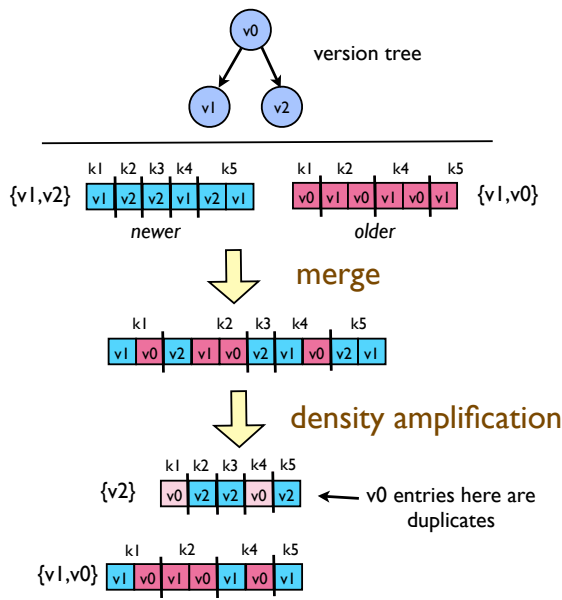

Figure 2: Simplified SDA process

**Density amplification.** The result of merging arrays together following promotion may be a large array which is not dense for all versions. For example, it may be tagged with two versions $v, w$ and have lots of elements for $v$ but too few for $w$, all potentially spread randomly through the array. To restore density we replace the large array with several smaller arrays each of which is dense – a process we call 'density amplification'. Crucially, by carefully choosing which sets of versions to include in which array, we are able to bound the amount of work done during density amplification, and to guarantee that not too many additional duplicate elements are created.

The basic density amplification process on an array $(A, W)$ operates by repeatedly searching for sets of versions that can be extracted together into a new dense subarray. In particular we are interested in subtrees – for a given version $v$, the subtree $W[v]$ is the set of all versions in $W$ that are descendant from $v$ (and $v$ itself) – and even more specifically, in cases where the array produced by extracting such a set of versions would be dense.

Roughly speaking the density amplification process searches for sets of sibling versions $v_i$ (i.e. with the same parent) whose version subtrees $W[v_i]$ can be amalgamated without destroying the density property. Once such a set $U = \cup_i W[v_i]$ is identified, the corresponding array is extracted, and the same process applied recur-sively to the remainder $(A, W \setminus U)$ (the precise notions are a bit more involved, see [6]).

**Lookups.** In the simplest construction, we embed into each array a B-tree and maintain a Bloom filter on the set of keys in each array. A lookup for version $v$ involves querying the Bloom filter for each array tagged with version $v$, then performing a B-tree walk on those arrays matching the filter. Typically, the internal nodes of the B-trees can be held in memory so that this involves $O(1)$ IOs per array.

**Range queries.** The density property implies that, for large range queries, simply scanning the arrays with the appropriate version tags will use an asymptotically optimal number of IOs. For small range queries, we resort to a more involved 'local density' argument to show that, over a large enough number of range queries, the average range query cost will also be optimal. More details can be found in [6].

**Fractional cascading.** In a more involved version of the construction, elements from higher levels are sampled in order that lookups in lower levels need only examine small portions of higher-level arrays. These 'forward pointers' are sampled carefully as the arrays are constructed, and then merged into the lower-level arrays during promotions. Queries are then performed for a given version $v$ by searching inside all arrays at level 0 tagged with $v$, then following the forward pointers to locations in higher-level arrays.

**Static optimality** Stratified B-trees offer *static optimality*: that is, imagine you have a sequence of inserts and lookups, where element $x_j$ of version $v$ is requested with probability $p_{j,v}$. If you knew all the $p_{j,v}$ in advance, you could design a data structure optimized for this access pattern. By modifying themselves on-the-fly (similarly to splay trees [16]), stratified B-trees allow users to store lots of versioned data (they have high ingest rates), yet combined with static optimality and SSDs, offer extremely efficient (provably optimal) queries for unknown access patterns.

## 6.2 Block allocation and low free space

When a new array of size $k$ needs to be written, we try to allocate it in the first free region of size $\geq k$. If this fails, we use a 'chunking strategy': we divide each array into contiguous *chunks* of size $c >> B$ (currently 10MB). During a merge, we read chunks from each input array until we have a chunk formed in memory to output. When a chunk is no longer needed from the input chunk arrays, it can be deallocated and the output chunk written there. This doesn't guarantee that the the entire array is sequential on disk, but it is sequential to within the chunk size, which is sufficient in practice to extract good performance, and only uses $O(1)$ extra space during merges - thus the system degrades gracefully under low free space conditions.

## 7 Practicalities

A data structure can be theoretically good, but unless it is easy to implement efficiently, it won't be practical. Here we consider some of the major practical issues that face B-trees, and describe how, in many cases, the SDA in fact alleviates them. alleviate some of these problems, then these are of practical interest.

### 7.1 Cache-obliviousness and block sizes

Storage devices are complex, and so too is the interaction with complex memory hierarchies. The basic premise of cache-oblivious algorithmics [9] is that the traditional model of assuming a single, fixed block size $B$ is no longer a good approximation for interacting with storage devices. Stratified B-trees don't need block-size tuning, unlike B-trees. One major advantage is that they are naturally good candidates for SSDs, where a large asymmetry in read/write block sizes makes life very hard for B-trees. In contrast, the arrays making up the SDA can be written with extremely large effective block sizes.

### 7.2 Consistency

We would like the data structure to be *consistent* – at any time, the on-disk representation is in a well-defined and valid state. In particular, we'd like it to always be the case that, after a crash or power loss, the system can continue without any additional work (or very little) to recover from the crash.

Both CoW and SDA implement consistency in a similar way: the difference between the new and old data structure is assembled in such a way that it can be abandoned at any time up until a small in-memory change is made at the end. In the case of CoW it is the root for version $v$: until the table mapping $v$ to root node is updated, the newly copied nodes are not visible in searches, and can be recycled in the event of a crash. Likewise for the SDA arrays are not destroyed until they are no longer in the search path, thus ensuring consistency.

### 7.3 Concurrency

With many-core architectures, easily exploiting concurrency is crucial to achieving decent performance. Concurrency in B-trees is a well-studied problem, and is quite a difficult challenge, particularly in the presence of multiple writers - Rodeh [14] has a good discussion about lock-coupling and other techniques.

The SDA can naturally be implemented without locks. Each embedded B-tree is built bottom-up in a 'reverse DFS order' (i.e. a node is only written once all its children have been written), and once a node is written, it is immutable. If there is no writer, then clearly there is no contention among the readers. If the B-tree is being constructed as a merge of $T_1, T_2$, then there is a *partition key* $k$ such that all nodes with key less than $k$ are immutable. Searches for key $< k$ go to the immutable subtree $T_{<k}$ with no contention. Searches for key $\geq k$ go to the immutable trees $T_1, T_2$ (searching both of them concurrently). Each merge can thus be handled with a single writer without locks. In the SDA, there are multiple merges ongoing concurrently, and each such merge is handled by a separate writer.a there are often enough concurrent merges to fully utilise all available cores.

## 8 Experimental results

We implemented prototypes of the Stratified Doubling Array and CoW B-tree (using in-place updates) in OCaml. The machine had 1GB memory available, a 2GHz Athlon 64 processor (although our implementation was only single-threaded) and a 500GB SATA disk. We used a block size of 32KB; the disk can perform about 100 such IOs/s.

We started with a single root version and inserted random 100 byte key-value pairs to random leaf versions, and periodically performed range queries of size 1000 at a random version. Every 100,000 insertions, we create a new version as follows: with probability 1/3 we clone a random leaf version and w.p. 2/3 we clone a random internal node of the version tree. The aim is to keep to the version tree 'balanced' in the sense that there are roughly twice as many internal nodes as leaves.

The figures show results for the CoW (btree), the Stratified Doubling Array (strat-DA) and, for comparison, the SDA where we disable density amplification; hence there is a single array at each level after promotions. Figure 3 shows the insertion performance on the disk. The B-tree performance degrades rapidly when the index exceeds internal memory available. Figure 4 shows range query performance (elements/s extracted using range queries of size 1000) - the SDA beats the CoW B-tree by a factor of more than 10. Interestingly, the CoW B-tree is limited by random IO here[5], but the SDA is CPU-bound (OCaml is single-threaded). Preliminary performance results from a highly-concurrent in-kernel implementation suggest that over 500,000 versioned updates/s are possible with 16 cores.

---

[5](100/s*32KB)/(200 bytes/key) = 16384 key/s

## 9 Conclusions

We believe the stratified B-tree construction outlined here (and the more detailed construction described in [6]) represents a significant step forward in the story of versioned data structures. We are currently developing an industrial-strength (but open-source) implementation in the Linux kernel, performing more detailed tests and developing improved techniques for, e.g. deleting stale versions.

## References

[1] R. Bayer and E. McCreight. Organization and maintenance of large ordered indexes. *Acta Informatica*, 1(3):173–189, 1972.

[2] Bruno Becker and Stephan Gschwind. An asymptotically optimal multiversion b-tree. *The VLDB Journal*, 5(4):264–275, 1996.

[3] Michael A. Bender and Martin Farach-Colton et al. Cache-oblivious streaming b-trees. In *SPAA '07*, pages 81–92, New York, NY, USA, 2007. ACM.

[4] Jeff Bonwick and Matt Ahrens. The zettabyte file system, 2008.

[5] Gerth Stolting Brodal and Rolf Fagerberg. Lower bounds for external memory dictionaries. In *SODA '03*, pages 546–554, Philadelphia, PA, USA, 2003. Society for Industrial and Applied Mathematics.

[6] A. Byde and A. Twigg. Optimal query/update tradeoffs in versioned dictionaries. http://arxiv.org/abs/1103.2566. *ArXiv e-prints*, March 2011.

[7] J R Driscoll and N Sarnak. Making data structures persistent. In *STOC '86*, pages 109–121, New York, NY, USA, 1986. ACM.

[8] Btrfs file system. http://en.wikipedia.org/wiki/Btrfs.

[9] Matteo Frigo and Charles Leiserson. Cache-oblivious algorithms. In *FOCS '99*, pages 285–, Washington, DC, USA, 1999. IEEE Computer Society.

[10] Dave Hitz and James Lau. File system design for an nfs file server appliance, 1994.

[11] Intel x25m g2 ssd data sheet, ftp://download.intel.com/newsroom/kits/ssd/pdfs/X25-M_34nm_DataSheet.pdf, 2011.

[12] Sitaram Lanka and Eric Mays. Fully persistent B+-trees. *SIGMOD Rec.*, 20(2):426–435, 1991.

[13] John T. Robinson. Analysis of steady-state segment storage utilizations in a log-structured file system with least-utilized segment cleaning. *SIGOPS Oper. Syst. Rev.*, 30:29–32, October 1996.

[14] Ohad Rodeh. B-trees, shadowing, and clones. *Trans. Storage*, 3:2:1–2:27, February 2008.

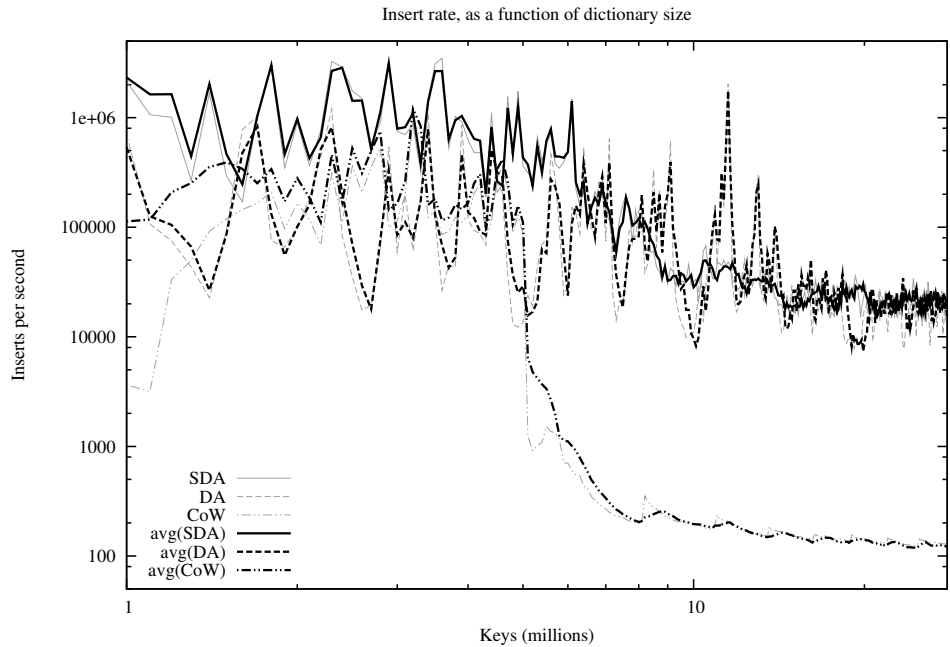

Figure 3: insert performance with 1000 versions.

[15] Mendel Rosenblum and John K. Ousterhout. The design and implementation of a log-structured file system. *ACM Trans. Comput. Syst.*, 10:26–52, February 1992.

[16] Daniel Dominic Sleator and Robert Endre Tarjan. Self-adjusting binary search trees. *J. ACM*, 32:652–686, July 1985.

[17] Craig A. N. Soules, Garth R. Goodson, John D. Strunk, and Gregory R. Ganger. Metadata efficiency in versioning file systems. In *Proceedings of the 2nd USENIX Conference on File and Storage Technologies*, pages 43–58, Berkeley, CA, USA, 2003. USENIX Association.

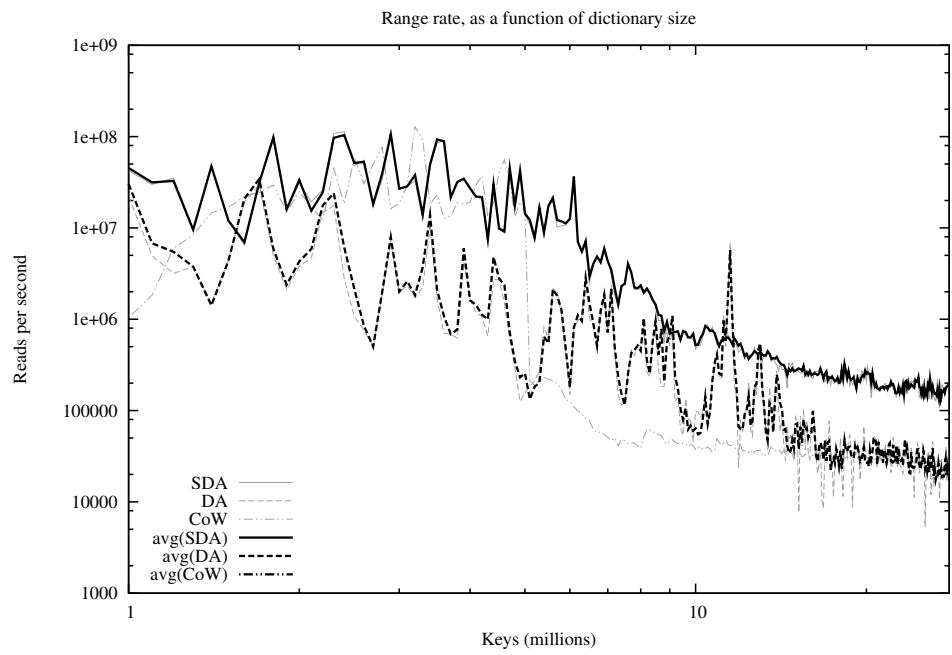

Figure 4: range query performance with 1000 versions.